\documentclass[%
 reprint,
superscriptaddress,
 amsmath,amssymb,
 aps,
]{revtex4-2}

\usepackage{graphicx}
\usepackage{dcolumn}
\usepackage{bm}
\usepackage{siunitx}
\usepackage{comment}

\begin{document}

\preprint{APS/123-QED}
\title{Correlated Dephasing in a Piezoelectrically Transduced Silicon Phononic Waveguide}

\author{Oliver A. Hitchcock}
 \email{oliver.hitchcock@stanford.edu}
 \affiliation{Department of Physics, Stanford University, Stanford, California 94305, USA}
  \affiliation{\mbox{Department of Applied Physics and Ginzton Laboratory, Stanford University, Stanford, California 94305, USA} }
\author{Felix~M.~Mayor}
 \affiliation{\mbox{Department of Applied Physics and Ginzton Laboratory, Stanford University, Stanford, California 94305, USA} }
\author{Wentao~Jiang}
\author{Matthew~P.~Maksymowych}
\author{Sultan~Malik}
\author{Amir~H.~Safavi-Naeini}
\email{safavi@stanford.edu}
\affiliation{\mbox{Department of Applied Physics and Ginzton Laboratory, Stanford University, Stanford, California 94305, USA} }
\date{\today}
\begin{abstract}
Nanomechanical waveguides offer a multitude of applications in quantum and classical technologies. Here, we design, fabricate, and characterize a compact silicon single-mode phononic waveguide actuated by a thin-film lithium niobate piezoelectric element. Our device directly transduces between microwave frequency photons and phonons propagating in the silicon waveguide, providing a route for coupling to superconducting circuits. We probe the device at millikelvin temperatures through a superconducting microwave resonant matching cavity to reveal harmonics of the silicon waveguide and extract a piezoelectric coupling rate $g/2\pi=\SI{1.1}{\MHz}$ and a mechanical coupling rate $f/2\pi=\SI{5}{\MHz}$. Through time-domain measurements of the silicon mechanical modes, we observe energy relaxation timescales of $T_{1,\text{in}} \approx \SI{500}{\micro\second}$, pure dephasing timescales of $T_\phi \approx \SI{60}{\micro\second}$ and dephasing dynamics that indicate the presence of an underlying frequency noise process with a non-uniform spectral distribution. We measure phase noise cross-correlations between silicon mechanical modes and observe detuning-dependent positively-correlated frequency fluctuations. Our measurements provide valuable insights into the dynamics and decoherence characteristics of hybrid piezoelectric-silicon acoustic devices, and suggest approaches for mitigating and circumventing noise processes for emerging quantum acoustic systems.
\end{abstract}
\maketitle

\section{\label{sec:Intro}Introduction}
Guided waves of light and sound are ubiquitous in modern technology. 
Compared to electromagnetic waves, acoustic waves propagate at much lower velocities, enabling miniaturization of gigahertz frequency devices and offering opportunities for on-chip integration~\cite{Gad2001,Roukes2005,Cleland1996}. Indeed, acoustic resonators and filters form the backbone of timekeeping and signal-processing technologies~\cite{Morgan2007,campbell2012,Delsing2019}, and can operate in the quantum regime when cooled to millikelvin temperatures~\cite{Manenti2016,O’Connell2010}. Coupling such gigahertz (GHz) phonons to superconducting circuits has already opened pathways toward hybrid quantum information processing and sensing~\cite{Chu2017,Chu2018,Bild2023,Satzinger2018,Cleland2024}. 
Realizing their full potential, however, demands waveguides that confine and selectively excite only the desired acoustic mode over a large bandwidth without scattering into spurious channels~\cite{Safavi2019}.  
While cavity optomechanics has been used to transduce modes of nanoscale phononic waveguides~\cite{Fang2016,Patel2018,Zivari2022}, its integration with microwave superconducting circuits remains a challenge that is yet to be addressed. In contrast to cavity optomechanics, direct piezoelectric or capacitive electrostatic actuation~\cite{Bozkurt2023} of phononic waveguides offers direct compatibility with integrated microwave circuits and emerging superconducting quantum machines~\cite{Chu2020}. However, each approach faces significant challenges. Electrostatic actuation requires small vacuum gaps and high bias voltages for efficient acoustic excitation, which can lead to mechanical instabilities~\cite{Zhao2023,bozkurt2024mechanicalquantummemorymicrowave}. Conversely, piezoelectric transduction is hindered by crystal defects~\cite{phillips_1987} and fabrication complexities that limit acoustic coherence times~\cite{Wollack2021,Gruenke2024}. Moreover, both methods rely on transducers with dimensions comparable to those of single-mode GHz acoustic waveguides. While these micron-scale transducers help mitigate scattering into spurious modes, they inherently possess very low capacitance resulting in significant impedance mismatches with conventional microwave circuits and leading to inefficient electrical excitation.

\begin{figure*}[]
    \centering
    \includegraphics[width=\linewidth]{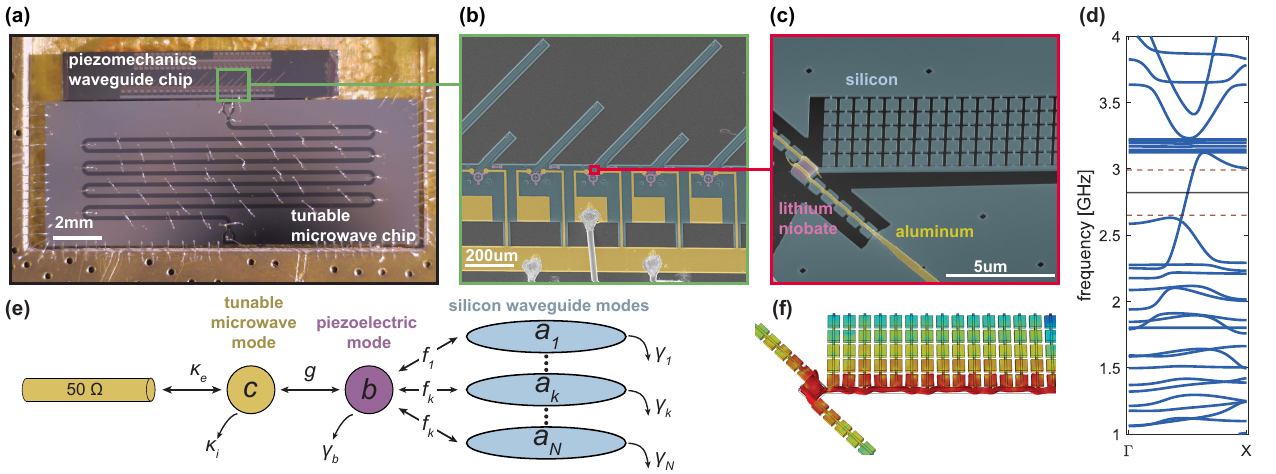}\caption{\label{fig:wide}\textbf{Piezoelectrically transduced silicon phononic waveguide}. (a) Optical microscope image showing the piezomechanics chip and the high-impedance microwave resonator chip. (b) A false color scanning electron micrograph (SEM) showing an array of piezoelectrically transduced waveguide devices, with the lithium niobate (purple), silicon (cyan) and aluminum electrodes (yellow). (c) An angled SEM showing a single waveguide device near the piezoelectric transducer region. (d) Simulated mechanical bands of the single-mode waveguide. The single-mode region is delimited by the two dashed lines. (e) Coupled mode model for the device. (f) Simulated mechanical mode profile of the single-mode waveguide.}
    \label{fig:fig1}
\end{figure*}

Here, we address these challenges by presenting a mechanical waveguide device that combines the strong piezoelectricity of thin-film lithium niobate~\cite{Weis1985} with the potentially long coherence times of silicon nano-acoustic resonators~\cite{MacCabe2020}. To leverage the respective advantages of these different materials, we heterogeneously integrate the lithium niobate with silicon through a transfer print technique~\cite{Meitl2006,Jiang2023}. We mitigate the impedance mismatch problem with the strong piezoelectricity of lithium niobate and resonant enhancement via an electromagnetic resonator. We then characterize the device with microwave spectroscopy to extract the relevant system parameters, and use time-domain ringdowns to measure decoherence and correlated dephasing dynamics. 

The silicon waveguide is engineered similarly to that presented in Ref.~\cite{Patel2018}. A one-dimensional (1D) phononic crystal with a full bandgap is terminated on one side to realize a 1D edge defect. Proper design of the defect can realize a single-mode phononic crystal that eliminates scattering to spurious mechanical polarizations within the waveguide. This design results in a waveguide that supports only longitudinal modes with group velocities $v_g=\SI{5200}{\meter\per\second}$ between frequencies of $\qtyrange{2.6}{3.0}{\GHz}$. By terminating the waveguide at a length of $L=\SI{700}{\micro\meter}$, we produce standing waves with a free spectral range (FSR) of $\omega_{\text{FSR}}/2\pi \approx \SI{3.7}{\MHz}$. The simulated mechanical bands of this single-mode waveguide are shown in Fig.~\ref{fig:fig1}(d).
We heterogeneously integrate a block of thin-film lithium niobate with an area of approximately $\SI{1}{\um\squared}$ at one end of the silicon phononic waveguide. This block of lithium niobate serves as a transducer that directly converts between microwave frequency electromagnetic and mechanical excitations. We design the frequency of the lithium niobate transducer mode $b$ to be $\omega_b/2\pi=\SI{2.8}{\GHz}$, coinciding with the single-mode region of the silicon phononic waveguide. Aluminum electrodes provide the electrical interface with the piezoelectric block. The micron-sized block has very little capacitance $C_k\approx\SI{0.2}{\femto\farad}$ and is poorly matched to a $~\SI{50}{\ohm}$ transmission line. We therefore connect the aluminum electrodes via wirebond to a tunable high-impedance ($Z\approx\SI{500}{\ohm}$) microwave resonator on a separate chip [Fig.~\ref{fig:fig1}(a)]. The microwave resonator compensates for the impedance mismatch between the transducer and the transmission lines, and is made of a niobium-titanium-nitride (NbTiN) ladder structure which allows for in-situ tuning of the frequency via an external magnetic coil~\cite{Han2020, Malik2023}. The lithium niobate transducer section of the device is shown in Fig.~\ref{fig:fig1}(c) and the simulated displacement of the transducer is shown in Fig.~\ref{fig:fig1}(f). An overview showing the extent of the waveguide is shown in Fig.~\ref{fig:fig1}(b).

\begin{figure*}[]
    \centering
    \includegraphics[width=\linewidth]{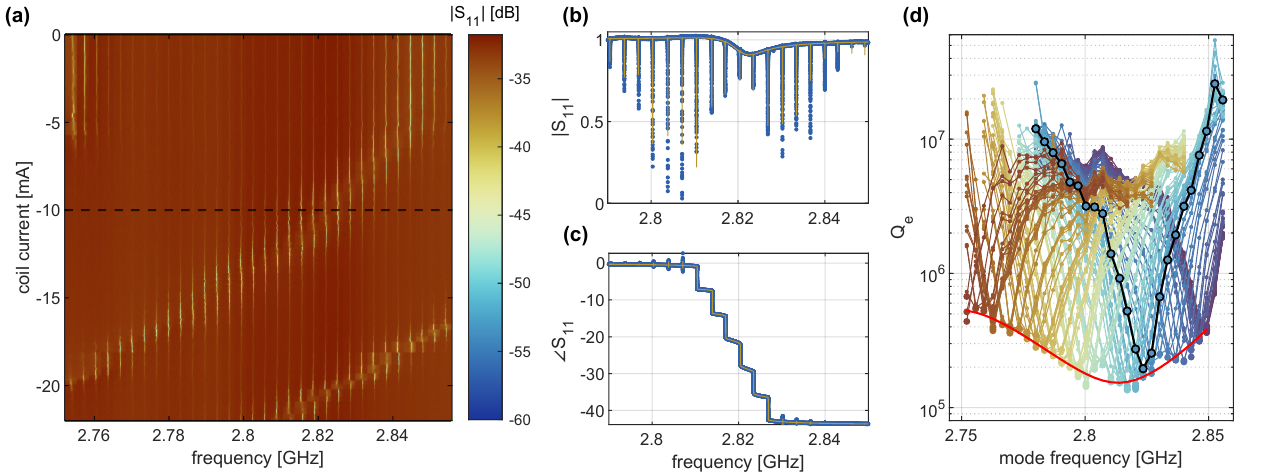}
    \caption{\label{fig:wide}\textbf{Device characterization}. (a) Reflection magnitude as the microwave mode frequency is tuned by the superconducting coil. Sharp lines correspond to resonances in the mechanical waveguide. (b) Reflection magnitude and (c) phase versus frequency at \SI{10}{\milli\kelvin} are shown with the background removed (blue) and fit to the coupled mode model (yellow). (d) Coupling enhancement as a function of mode frequency as the microwave mode is tuned. Each colored curve represents the fit $Q_\text{e}$'s for mechanical modes at a different current in the tuning coil. Overlaying all curves reveals a region of enhanced coupling near the bare lithium niobate mode frequency $\omega_b/2\pi\approx 2.81$ GHz. A fit to the minimum $Q_\text{e}$ value for each coil current is shown in red.}
    \label{fig:fig2}
\end{figure*}

We model the device using coupled mode theory as represented schematically in Fig.~\ref{fig:fig1}(e). The terminated silicon waveguide gives rise to a series of standing-wave modes $a_k$ with frequencies $\omega_{a,k}$ and intrinsic loss rates $\gamma_{i,k}$. These modes are mechanically coupled at rates $f_k$ to the lithium niobate piezoelectric mode $b$ and the interaction leads to a collection of beam splitter coupling terms, $H_{\text{mech}}/\hbar=\sum f_k (b^\dagger a_k +\text{h.c.})$. The lithium niobate mode $b$ at a frequency $\omega_b$ has an intrinsic loss rate of $\gamma_b$ and is piezoelectrically coupled at rate $g$ to the microwave mode ${c}$ with frequency $\omega_c$, giving rise to a piezoelectric coupling term,  $H_{\text{pe}}/\hbar= g ({c}^\dagger{b} + \text{h.c.})$. The full Hamiltonian is,
\begin{equation}\label{eq:full_hamiltonian}
    H/\hbar = \omega_c c^\dagger c+\omega_b b^\dagger b + \sum_k \omega_{a,k}{a}_k^\dagger {a}_k + {H}_{\text{pe}}+ {H}_{\text{mech}},
\end{equation}
from which we  derive the one-port scattering parameter $S_{11}$, which describes the reflection coefficient when the system is probed at the microwave mode:
\begin{equation}\label{eq:full_S11}
S_{11}(\omega) =     1-\frac{\kappa_e}{\chi_c^{-1} + g^2/(\chi_b^{-1} - i\sum_k f_k^2 \chi_{a,k})}.
\end{equation}
Here, $\chi_c^{-1} = -i(\omega - \omega_c) + (\kappa_e+\kappa_i)/2$, $\chi_b^{-1} = -i(\omega - \omega_b) + (\gamma_b)/2$ , and $\chi_{a,k}^{-1} = -i(\omega - \omega_{a,k}) + (\gamma_{i,k})/2$  are the inverses of the susceptibilities for the microwave, lithium niobate and silicon waveguide modes, respectively. These terms capture how each mode responds to the input frequency $\omega$. This model and the associated scattering parameter forms the basis of our analysis of the system.

\section{\label{sec:characterization}Device characterization} 
We cool the device to \SI{10}{\milli\kelvin} and probe it by sending in a microwave tone and measuring the reflected signal as a function of frequency. Simultaneously, we apply a DC current, swept from \qtyrange{0}{-20}{\mA}, to an off-chip coil placed directly above the sample (corresponding roughly to a magnetic field of \qtyrange{0}{0.9}{\milli\tesla}) to sweep the $c$ mode frequency. We observe a mode of the NbTiN resonator that is overcoupled to a transmission line with extrinsic linewidth $\kappa_e/2\pi = \SI{8 \pm 1}{\MHz}$, and tunes from \qtyrange{2.85}{2.75}{\GHz}.
As the microwave mode is tuned, we observe a series of narrow dips [Fig.~\ref{fig:fig2}(a)] that become more prominent as they are brought closer to resonance with the $c$ mode. The narrow dips correspond to the high-$Q$ standing wave modes, $a_k$, in the nanomechanical waveguide that are hybridized with the LN transducer.
The mechanical modes are spaced by $\omega_{\text{FSR}}/2\pi=\SI{3.3\pm 0.2}{\MHz}$ and have an enhanced reflection response when in the proximity of the microwave mode due to the resonant enhancement in their out-coupling rate (Appendix~\ref{appendix:coupling_enhancement_microwave_mode}). Fig. \ref{fig:fig2}(b,c) shows the magnitude and phase response at a coil current of $\SI{-10}{\mA}$ corresponding to the black dashed line in Fig. \ref{fig:fig2}(a). In addition to the broad shallow dip of the $c$ mode, we clearly see the narrow $a_k$ mechanical modes that become more strongly coupled to the output channel when near the broad microwave mode at $\omega_c/2\pi=\SI{2.82}{\GHz}$.

To observe the narrow features corresponding to the $a_k$ modes, they must be hybridized with the piezoelectric element $b$. However, we do not observe  any feature that corresponds unambiguously with the $b$ mode as the device is in a regime where the mechanical coupling exceeds the FSR, $\gamma_{\text{WG}}>\omega_{\text{FSR}}$. In this regime, many waveguide modes are simultaneously strongly coupled to the $b$ mode \cite{Sundaresan2015}. In principle, a shift in the frequencies of the $a_k$ modes can be used to determine the frequency of the $b$ mode, however this dispersive effect is masked by the disorder in the $a_k$ mode frequencies of our device. 
Instead of looking for a dispersive shift in the $a_k$ mode frequencies, we try to measure the dissipation induced by coupling the $a_k$ modes to the lossy microwave mode through the $b$ mode. We sweep the frequency of the microwave mode $c$, and measure the extrinsic quality factor $Q_\text{e}$ of the hybridized mechanical modes to obtain $Q_\text{e}$ versus $a_k$ mode frequency for each $c$ mode frequency. In this measurement, we expect to see a reduction in $Q_\text{e}$ for $a_k$ modes close to the microwave resonance. We show the resulting $Q_\text{e}$ versus $a_{k}$ frequency for different coil currents (leading to differing $\omega_c$) as the different colored curves in Fig.~\ref{fig:fig2}(d). The $Q_\text{e}$'s decrease as the mode frequency is brought closer to $\omega_c$ and the out-coupling is resonantly enhanced (Appendix~\ref{appendix:coupling_enhancement_microwave_mode}). We see that the location of the dip for each curve corresponds to the $c$ mode frequency. By selecting the minimal $Q_\text{e}$ for each curve, we observe the envelope of minimum $Q_\text{e}$ which is itself minimized at a certain frequency. This is due to an additional resonant enhancement in the extrinsic coupling rate that results from the proximity of silicon waveguide mode frequencies to the bare lithium niobate mode frequency $\omega_b$. 
We fit this curve of increased out-coupling to a Lorentzian [red curve in Fig. \ref{fig:fig2}(d)], which yields a value of $\omega_b/2\pi=\SI{2.814}{\GHz}$ and agrees with fits to the diagonalized Hamiltonian of the coupled mode model (Appendix~\ref{appendix:fit_Qe_envelope_diagonalize_hamiltonian}). The linewidth of this Lorentzian, $\gamma_{\text{WG}}/2\pi=\SI{48.8}{\MHz}$, corresponds to the loss rate of the LN mode into the silicon waveguide, had the silicon waveguide been extended out to infinite length, \textit{i.e.}, the continuum limit for the mode spectrum. This coupling rate is a property of the mechanical coupling at the LN transducer-waveguide interface, independent of the length of the waveguide.

To determine the coupling rates $g$ and $f_k$, we fit the magnitude and phase of the reflection spectrum to the coupled mode model [Eq.~(\ref{eq:full_S11})]. For simplicity, we assume a constant $f_k$ across all modes. We use our model to fit spectra taken at different microwave frequencies $\omega_c$ using otherwise common parameters for the fits and use a range of initial guesses to estimate the uncertainties in the parameters. An example fit result overlaid with one spectrum is shown in Fig.~\ref{fig:fig2}(b). Our approach yields $\bar{f}_k/2\pi= \SI{5.0\pm 0.2}{\MHz}$ and $g/2\pi=\SI{1.1\pm 0.1}{\MHz}$. Notably, the fit value of $\bar{f}_k$ is in good agreement with the prediction $f_k=\sqrt{\omega_\text{FSR}\gamma_\text{WG}/2\pi}\approx \SI{5.1}{\MHz}$ when using the measured FSR and fit $\gamma_\text{WG}$ from the $Q_\text{e}$ envelope analysis.

\section{\label{sec:decoherence_and_correlated_dephasing}Decoherence and Correlated Dephasing}
To characterize and disentangle the effects of loss and dephasing in the mechanical waveguide, we probe the device in time domain. This allows us to separate the energy relaxation from the pure dephasing for each mode and to investigate correlated frequency fluctuations between mechanical modes in the waveguide, providing insight into the underlying microscopic processes.

Consider the evolution of a modal field amplitude undergoing energy dissipation at rate $\kappa$ and frequency fluctuations $\delta \omega(t)$ leading to a phase $\phi(t)=\int_t \delta\omega(t') dt'$,
\begin{equation}\label{eq:field_ringdown_evolution}
    a(t) = a_0 e^{-\kappa t/2} e^{-i\phi(t)}.
\end{equation}
If we sample the \textit{magnitude} of the cavity field amplitude during the ringdown we obtain the energy ringdown which depends only on the total energy decay rate,
\begin{equation}\label{eq:energy_decay_general}
    \langle |a(t)|^2\rangle = |a_0|^2e^{-\kappa t}.
\end{equation}
On the other hand, averaging the \textit{field} amplitude Eq.~(\ref{eq:field_ringdown_evolution}) before taking its magnitude, as is done in coherent spectroscopy experiments, leads to a different ring down,
\begin{equation}\label{eq:amplitude_decay_general}
    |\langle a(t)\rangle|^2 = |a_0|^2 e^{-\kappa t} |\langle e^{-i\phi(t)}\rangle|^2.
\end{equation}
The decay of Eq.~(\ref{eq:amplitude_decay_general}) depends on the details of the frequency noise process, as encapsulated in the stochastic function $\phi(t)$. But in general, $|\langle a(t)\rangle|^2$ will decay faster than $\langle |a(t)|^2\rangle$. Gaussian white frequency noise results in exponential decay of the average amplitude $|\langle a(t)\rangle|^2\sim e^{-(\kappa+\kappa_\phi)t}$, where $\kappa_\phi$ is the increase in the decay rate caused by frequency fluctuations. Other noise processes can give rise to different time dependencies, such as those quadratic in time (Appendix~\ref{appendix:frequency_phase_noise_models}).

\begin{figure}[]
    \centering
    \includegraphics[width=\linewidth]{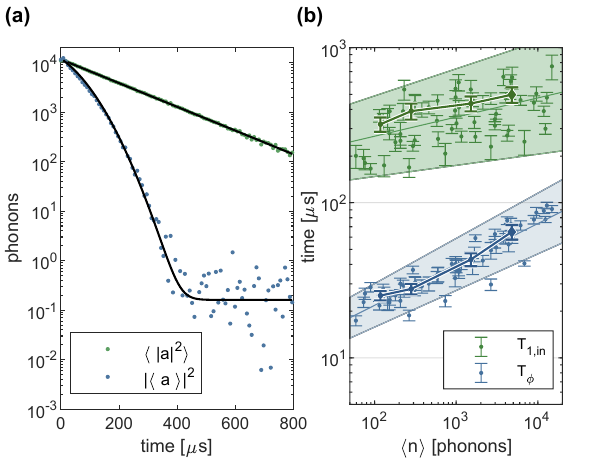}
    \caption{\label{fig:wide}\textbf{Energy and amplitude decay for individual modes}. (a) Ringdown of the cavity field magnitude (green) and cavity field amplitude (blue) with fits (black) for a waveguide mode shows energy decay is described by a simple exponential, whereas amplitude decay requires an additional exponential term that is quadratic in time. (b) Timescales of intrinsic energy decay $T_{1,\text{in}}$ and pure dephasing $T_\phi$ versus intracavity phonon number. Points represent individual modes probed at different RF drive powers, while error bars show the standard error of the mean from 24 repeated measurements. The mode measured in (a) is shown for different intracavity powers as the data points connected by lines in (b), with the highest intracavity population datapoint corresponding to the data from (a). Shaded regions in (b) correspond to 95\% confidence bounds on parameters for fits to power laws of $T_{1,\text{in}}\propto \langle n\rangle^{0.12}$ and $T_\phi\propto\langle n\rangle^{0.26}$}
    \label{fig:fig3}
\end{figure}
\begin{figure*}[]
    \centering
    \includegraphics[width=\linewidth]{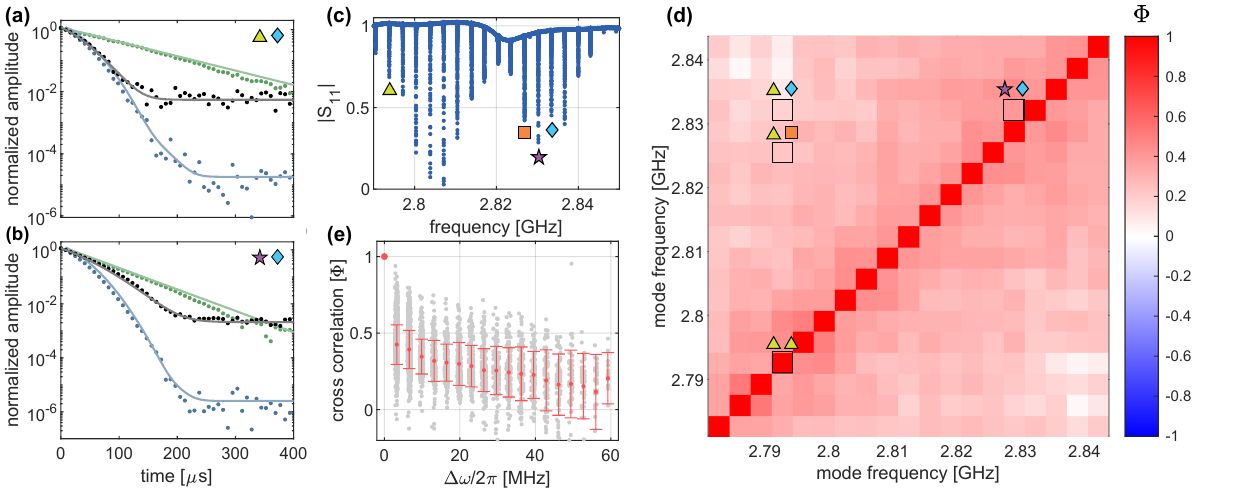}
    \caption{\label{fig:wide}\textbf{Equal-Time Cross-Correlation}. (a) Equal-time cross-correlation field ringdown (black) between far detuned modes at frequencies $\omega_1$ and $\omega_4$ ($\Delta\omega_{14}/2\pi=\SI{40}{\MHz}$) reveals predominantly uncorrelated phase noise when compared with the expected signals for perfectly correlated (uncorrelated) phase noise shown in green (blue). (b) Equal-time cross-correlation field ringdown between near detuned modes at frequencies $\omega_3$ and $\omega_4$ ($\Delta\omega_{34}/2\pi=\SI{4}{\MHz}$) reveals the presence of correlated phase noise. (c) Spectrum obtained with microwave mode frequency $\omega_c/2\pi\approx \SI{2.82}{\GHz}$, highlighting four waveguide modes at $\omega_1/2\pi=\SI{2.793}{\GHz}$ (yellow triangle), $\omega_2/2\pi=\SI{2.826}{\GHz}$ (orange square), $\omega_3/2\pi=\SI{2.829}{\GHz}$ (purple star) and $\omega_4/2\pi=\SI{2.833}{\GHz}$ (blue diamond).  (d) Matrix showing cross-correlation strengths $\Phi$ for each combination of waveguide modes. $\Phi$ is determined by comparing the fitted correlated dephasing timescale to the timescale associated with perfectly uncorrelated dephasing, averaged over experimental occurrence and nine microwave mode frequencies. (e) Equal-time cross-correlation strength averaged over detuning $\Delta \omega$ between modes.}
    \label{fig:fig4}
\end{figure*}
The energy and amplitude ringdown measurement results are shown as the green and blue points respectively in Fig.~\ref{fig:fig3}(a) for a mechanical mode at a frequency of $\omega_a/2\pi=\SI{2.7996}{\GHz}$ with the microwave mode detuned by approximately \SI{50}{\MHz}, where any dephasing induced by coupling to a noisy microwave mode is expected to be small. We fit the energy decay to a function of the form $A_0\exp(-\kappa t) + A_{\text{bg}}$, yielding the energy decay rate $\kappa$. The energy decay rate $\kappa$ contains both the extrinsic part due to coupling to microwaves $\kappa_e$ and the intrinsic part $1/T_{1,\text{in}}$. We use separate coherent spectroscopy measurements to obtain $\kappa_e$ which we subtract from the inferred $\kappa$ and get the resulting  $T_{1,\text{in}}$ [green points in Fig.~\ref{fig:fig3}(b)]. We fit the amplitude decay to a function of form $B_0\exp(-\kappa t - \Gamma t - \Delta^2 t^2) + B_{\text{bg}}$ with $\kappa$ fixed from the energy decay fit. The two decoherence rates $\Gamma$ and $\Delta$ arise from frequency fluctuation processes that are fast and slow respectively (Appendix~\ref{appendix:frequency_phase_noise_models}). We define our pure dephasing time scale $T_\phi$ as $(\Delta T_\phi)^2+\Gamma T_\phi=1$, \textit{i.e.}, the time after which coherences decay by $e^{-1}$ due to both slow and fast frequency noise [blue points in Fig.~\ref{fig:fig3}(b)]. The estimation of the intracavity phonon number associated with each measurement is detailed in Appendix~\ref{appendix:phonon_number_estimation_for_ringdown_measurements}. For the measurement shown in Fig.~\ref{fig:fig3}(a), we find $T_{1,\text{in}}=\SI{534}{\us}$, $T_{\phi}=\SI{107}{\us}$ and $\langle n\rangle\approx\SI{1e4}{}$. As evident from Fig.~\ref{fig:fig3}(a), the amplitude decay is much faster than the energy decay, suggesting $T_\phi$ that are significantly smaller than $T_{1,\text{in}}$. We fix the microwave mode frequency at $\omega_{c}/2\pi\approx\SI{2.85}{\GHz}$, and measure both time scales for 17 nearby modes as a function of incident power.  
We find that intrinsic energy relaxation timescales exceed dephasing by approximately a factor of 10. We also observe on average a weaker intracavity power dependence for the energy decay time, $T_{1,\text{in}}\propto \langle n\rangle^{0.12}$, as compared to the dephasing time, $T_\phi\propto\langle n\rangle^{0.26}$, both plotted in Fig.~\ref{fig:fig3}(b).

In addition to measurements on a single mode, we take measurements of products of field amplitudes, which tells us about correlations between frequency noise affecting different modes. Consider the equal-time cross-correlation which consists of products of amplitudes of different modes $a_j, a_k$:
\begin{equation}\label{eq:cross_corr_field}
    |\langle a_j(t) a_k^*(t)\rangle|  = |a_{j,0} a_{k,0}^*|e^{-\frac{\kappa_j+\kappa_k}{2}t}|\langle e^{-i[\phi_j(t) - \phi_k(t)]t }\rangle|.
\end{equation}
The last term in the expression is the difference in phase noise between the two modes. If both modes have precisely the same frequency fluctuations, \textit{i.e.}, perfectly correlated frequency noise, we would expect the argument of the exponential to be zero. We define a correlation strength metric $\Phi$ in Appendix~\ref{appendix:phase_noise_cross_correlation}, which for the perfectly correlated case would lead to $\Phi=1$. For the uncorrelated case, where the two modes have independent frequency or phase noise, the correlation strength would be given by $\Phi=0$. 

Our goal is to estimate $\Phi$ between different pairs of modes in our device through measurement. Fig.~\ref{fig:fig4}(c) shows an example spectrum with four modes highlighted by colored shapes for later reference.
Fig.~\ref{fig:fig4}(a) shows the equal-time cross-correlation (black) for two modes with large frequency difference indicated by the yellow triangle and blue diamond from Fig.~\ref{fig:fig4}(c). Fig.~\ref{fig:fig4}(b) shows the equal-time cross-correlation for two modes that are closer in frequency indicated by the purple star and blue diamond from Fig.~\ref{fig:fig4}(c). On the same plots, we also show the average of energy (amplitude) decay curves for the two modes in green (blue), which corresponds to a hypothetical perfectly correlated (uncorrelated) case of $\Phi=1$ ($\Phi=0$). We fit the cross-correlated data to the same functional form as the amplitude decay. In general, the equal-time cross-correlation signal decays more slowly than for uncorrelated dephasing, indicating some correlation in the phase noise between different waveguide modes. 

We measure and fit the cross-correlation signal for all pairs of modes and obtain the corresponding $\Phi$. In Fig.~\ref{fig:fig4}(d) we plot $\Phi$ as a function of a pair of mode indices for a collection of 19 modes. To obtain more data, we measure each $\Phi$ for nine different values of the microwave mode frequency and combine the results. Highlighted on this map are several mode pairs from Fig.~\ref{fig:fig4}(a,b,c). The diagonal of this map represents the equal-time self-correlation, which by definition is equal to $\Phi=1$. From this map, we see that near-detuned modes tend to have higher cross-correlation strengths than far-detuned modes. This trend becomes clearer when we average the cross-correlation strengths over detuning [red points in Fig.~\ref{fig:fig4}(e)], error bars indicate the standard deviation of all mode pair measurements [gray points in Fig.~\ref{fig:fig4}(e)] for that particular detuning. We find that the cross-correlation strengths have a large amount of variance, but on average diminish with increased detuning.

\section{\label{sec:conclusions}Conclusions}

We have measured decoherence and demonstrated evidence of correlated dephasing dynamics between different modes in a single-mode silicon acoustic waveguide. We engineered a heterogeneously integrated lithium niobate actuated silicon mechanical waveguide which combines the strong piezoelectricity of lithium niobate and the low losses of silicon. We characterize the device to find a piezoelectric coupling $g/2\pi=\SI{1.1}{\MHz}$ and mechanical coupling $f/2\pi=\SI{5}{\MHz}$. For different modes in the system, we measure energy decay  $\langle|a|\rangle^2$ to extract intrinsic decay timescales exceeding \SI{100}{\us} and amplitude decay $|\langle a \rangle|^2$ to extract pure dephasing timescales exceeding $\SI{10}{\us}$. We measure products of field amplitudes to determine equal-time cross-correlation strengths between different modes and find evidence of correlated frequency noise with strengths that depends on the relative detuning between the modes. 

Our results introduce a new device to the toolbox of phononic circuit elements that allows them to interface with microwave superconducting circuits, with potential uses in memory, information routing and sensing. The frequencies of the device can be engineered between $\sim\SI{100}{\MHz}$ and $\sim\SI{5}{\GHz}$ to suit one's application with reasonable changes to the silicon waveguide and lithium niobate dimensions. The FSR of the device can be arbitrarily modified by designing the length of the waveguide, limited only by the physical size of the chip. The piezoelectric coupling rate can be increased through better impedance matching by switching to flip-chip packaging, increasing the volume of piezoelectric that experiences an electric field, or increasing the impedance of the external microwave circuit. The mechanical coupling rate $\gamma_{\text{WG}}$ is determined by modal overlap at the interface between the transducer and waveguide regions of the device, therefore, this interface can be optimized through modal engineering to provide a path toward increased mechanical coupling. 

The main limitations of our current device, which will be the subject of future work, are the energy and phase coherence times, and the coupling rates. Assuming conservative values for low phonon occupations based on our measurements, $T_{1,\text{in}}=\SI{100}{\us}$, $T_{\phi}=\SI{10}{\us}$, would imply intrinsic energy and pure dephasing attenuation per length of \SI{0.7}{\dB\per\meter} and \SI{7}{\dB\per\meter} respectively. We suspect that a combination of resonator design, fabrication imperfections and near-resonant two-level-system (TLS) defects contribute to energy loss, whereas frequency noise is primarily the result of coupling to an ensemble of TLS defects that are distributed in frequency~\cite{maksymowych2025}. TLS as a source of frequency noise is consistent with our observations of equal-time cross-correlated dephasing between different modes. For modes coupled to a common bath of TLS, near-detuned modes would be predominantly influenced by a single TLS and are therefore correlated, whereas far-detuned modes are influenced by different TLS whose fluctuations are uncorrelated. This result is consistent with work on surface acoustic wave resonators measured over long timescales~\cite{Tubsrinuan2024}. Furthermore, there is considerable evidence that TLS defects are present on the surfaces and material interfaces of devices~\cite{MacCabe2020,Chen2024, gruenkefreudenstein2025}. Careful choice of materials and improved fabrication techniques, in particular the removal of silicon native oxide, will reduce the density of these defects along with the associated loss and dephasing they induce. 

Importantly, we note that while reducing the sources of frequency noise in the system is advantageous for the coherence of individual modes, the non-Markovian properties of the frequency noise open opportunities for techniques that extend the coherence times, such as dynamical decoupling~\cite{Bylander2011,miao2020}. Meanwhile, \textit{strong} frequency noise correlations may also prove useful for extending coherence times by encoding information in multiple modes, e.g. dual-rail or delay line schemes~\cite{Wan2021}. 
Further optimization of the device parameters, and operation as a delay line and integration with superconducting qubits or other nonlinear circuits such as asymmetrically threaded squids~\cite{Lescanne2020, Makihara2024} for circuit quantum acoustodynamics will be explored in future work.
\begin{acknowledgments}
We thank T. Makihara, K.K.Ss Multani, C.J. Sarabalis, R.N. Patel, M. Roukes, M. Dykman, M. Yuksel, and S. Gyger for helpful discussions, K.A. Villegas Rosales at Quantum Machines for technical support.
The authors gratefully acknowledge support from Amazon Web Services Inc., the Air Force Office of Scientific Research and Office of Naval Research Phononics MURI program (award FA9550-23-1-0338), the National Science Foundation CAREER Program (award ECCS-1941826), and the U.S. Army Research Office (ARO)/Laboratory for Physical Sciences (LPS) through the Modular Quantum Gates (ModQ) program (Grant W911NF-23-1-0254). This work was also supported by the U.S. Department of Energy through grant DE-AC02-76SF00515 and the Q-NEXT Center. Additional support was provided by a Moore Inventor Fellowship.
Part of this work was performed at the Stanford Nano Shared Facilities (SNSF) and at the Stanford Nanofabrication Facility (SNF), supported by the National Science Foundation under award ECCS-2026822.

\end{acknowledgments}


\appendix
\renewcommand{\figurename}{Supplementary Figure}
\setcounter{figure}{0}

\section{\label{appendix:device_fabrication}Device fabrication}
The fabrication of the device relies on hybrid integration of thin film lithium niobate and thin film silicon via the same transfer printing technique used for microwave to optical quantum frequency transduction devices \cite{Jiang2023}. We start with a lithium niobate (LN) on silicon dioxide insulator (LNOI) chip. The LN transducer is patterned with hydrogen silsequioxane (HSQ) and a 100kV electron beam lithography (EBL) system (JEOL JBX-6300FS) and etched by argon ion milling (Intlvac Nanoquest). The \SI{250}{\nano\meter} thick LN is cleaned and partially released in hydrofluoric acid (HF) before it is transferred onto a piranha cleaned \SI{220}{\nano\meter} silicon on insulator (SOI) chip using a PDMS stamping technique~\cite{Meitl2006}. The transferred sample is annealed at \SI{500}{\celsius}, then cleaned with a piranha solution. The silicon waveguide is patterned with EBL and etched in a $\text{Cl}_2/\text{HBr}$ chemistry (Lam Research TCP 9400). The device electrodes are patterned with EBL, whereas the large wirebond contacts are defined via photolithography (Heidelberg MLA150). Each metal deposition involves evaporating aluminum (Plassys MEB550S) followed by liftoff in \SI{80}{\celsius} Remover PG. The final step of the fabrication involves releasing the structure in anhydrous HF (SPTS uetch). The finished devices are glued into a PCB along with the separately fabricated microwave resonator chip \cite{Malik2023,Jiang2023} and connected by cross-chip wirebonds (West Bond 7476E).

\section{Device parameters}\label{appendix:device_parameters}
Table~\ref{table:device_parameters} presents a summary of several key device parameters obtained through spectroscopic microwave measurements of the device and fitting to the coupled mode model.
\begin{table}[h!]\centering
\begin{tabular}{c  c  c} 
 \hline \\ [-1.5ex] 
 Parameter   & Value & Description\\ [1.0ex] 
 \hline \\ [-1.5ex]
$\omega_c/2\pi$ & \qtyrange{2.75}{2.9}{\GHz} & microwave frequency \\ [0.5ex] 
$\kappa_e/2\pi$ & \SI{8\pm1}{\MHz} & microwave extrinsic loss \\[0.5ex] 
$\kappa_i/2\pi$ & \SI{250\pm100}{\kHz} & microwave intrinsic loss \\[0.5ex] 
$g/2\pi$        & \SI{1.1\pm0.1}{\MHz}  & piezoelectric coupling \\[0.5ex]
$\omega_b/2\pi$ & \SI{2.81\pm 0.05}{\GHz} & lithium niobate frequency\\[.5ex] 
$\gamma_b/2\pi$ & \SI{10\pm 5}{\kHz}  & lithium niobate loss\\[0.5ex] 
$\gamma_{\text{WG}}/2\pi$  & \SI{47\pm 2}{\MHz} & mechanical waveguide coupling\\ [1ex]
\hline
\end{tabular}
\caption{Device parameters}
\label{table:device_parameters}
\end{table}
\section{Simulated device and equivalent circuit }\label{appendix:simulated_device_parameters}
\begin{figure}
    \centering
    \includegraphics[width=\linewidth]{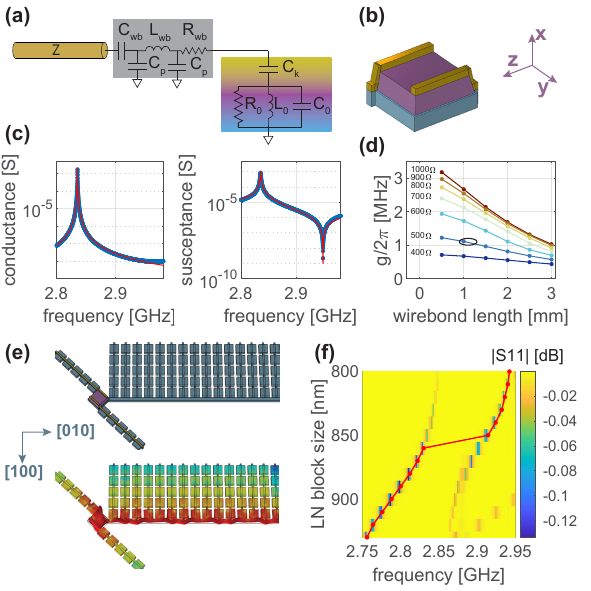}
    \caption{\textbf{Simulated piezoelectric and mechanical coupling rates}. (a) Equivalent circuit model for a high impedance microwave transmission line coupled via a wirebond to the piezoelectric transducer. (b) COMSOL transducer geometry with aluminum (yellow), lithium niobate (purple) and silicon (blue) as well as the LN crystal axes orientation. (c) Simulated admittance and fit to the transducer equivalent circuit. (d) Piezoelectric coupling versus wirebond length and microwave transmission line impedance, compared to the measured device's $g$ and approximate wirebond length (shaded grey region). (e) Simulated piezoelectric waveguide geometry and mode displacement profile. (f) Simulated $S_{11}$ as the piezoelectric block size is swept, showing an avoided crossing with the waveguide.}
    \label{fig:fig_App_SimulatedDeviceParams}
\end{figure}

We simulate the piezoelectric coupling rate, $g$, and the mechanical coupling rate, $f$, using finite element method (FEM) in COMSOL Multiphysics as part of the design process and for comparison with measurements. To simulate the piezoelectric coupling rate, $g$, we begin with a simplified geometry including only the transducer section of the device [Fig.~\ref{fig:fig_App_SimulatedDeviceParams}(b)], comprising a lithium niobate block (purple) on silicon (blue) with aluminum electrodes (yellow) and set an arbitrary material loss. We model this section of the device as a lossy LC circuit described by an admittance,
\begin{equation}
     Y(\omega) = \left[\frac{1}{i\omega C_k}+\left[i\omega C_0 + \frac{1}{i\omega L_0} + \frac{1}{R_0}\right]^{-1}\right]^{-1}.
\end{equation}
We simulate the low frequency response of the structure to estimate the coupling capacitance and find $C_k\approx \SI{0.2}{\femto\farad}$. We then use the frequencies of the zero and pole in the admittance caused by the resonance [blue points in Fig.~\ref{fig:fig_App_SimulatedDeviceParams}(c)] to estimate the equivalent circuit parameters $C_0, L_0$. We insert these estimates as the initial guess for a least squares fit to the full admittance [red curve in Fig.~\ref{fig:fig_App_SimulatedDeviceParams}(c)] and obtain values of $C_k=\SI{0.224}{\femto\farad}, C_0=\SI{2.76}{\femto\farad}, L_0=\SI{1.056}{\micro\henry}, R_0=\SI{100}{\mega\ohm}$. This implies a characteristic impedance for the piezomechanics equivalent circuit of $Z_0=\SI{19.5}{\kilo\ohm}$. We then construct a full circuit model [Fig.~\ref{fig:fig_App_SimulatedDeviceParams}(a)] that incorporates the wirebond and the microwave mode idealized as a high-impedance transmission line. As the length of the wirebond and the impedance of the microwave transmission line is swept, we calculate the resulting piezoelectric coupling rate $g$ [Fig.~\ref{fig:fig_App_SimulatedDeviceParams}(d)]. From optical images, we estimate the length of the wirebond to be $L_{\text{wb}}\approx \SI{1.1}{\mm}$ and from fits to the microwave mode spectra we find the characteristic impedance of the microwave mode to be $Z\approx\SI{500\pm100}{\ohm}$. These values predict a piezoelectric coupling rate of $g_{\text{sim}}/2\pi\approx \SI{1.1}{\MHz}$, which agrees with fits to measured data using the full coupled mode model.

To simulate the mechanical coupling rate, $f$, we combine the transducer section [Fig.~\ref{fig:fig_App_SimulatedDeviceParams}(b)] with a truncated (15 periods) silicon single-mode waveguide geometry, shown in Fig.~\ref{fig:fig_App_SimulatedDeviceParams}(e). The waveguide is separately optimized in the infinite extent limit to have a large fractional bandgap at the frequencies of interest. By sweeping the dimensions of the transducer (tuning the transducer frequency) and simulating the admittance versus frequency, we calculate $|S_{11}|$ versus transducer frequency and observe an avoided crossing between the transducer and waveguide mode [Fig.~\ref{fig:fig_App_SimulatedDeviceParams}(f)], allowing us to extract a mechanical coupling rate which can then be scaled to the waveguide length corresponding to our device, $f_{\text{sim}}/2\pi=\SI{5.25}{\MHz}$, roughly agreeing with the measured data.

\section{Fitting the $Q_\text{e}$ envelope to the diagonalized Hamiltonian}\label{appendix:fit_Qe_envelope_diagonalize_hamiltonian}

The Hamiltonian for the system can be expressed in matrix form as,
\begin{align}\label{eq:matrix_form_hamiltonian}
     H =
    \begin{pmatrix} \tilde\omega_c& ig & 0 & 0 & 0 &  \cdots \\
    ig&\tilde\omega_b& if & if &if& \cdots\\
    0&if& \tilde\omega_{a,1} & 0 & 0 &  \\ 
    0&if& 0 & \tilde\omega_{a,2} &0 & \\
    0&if& 0 & 0 & \tilde\omega_{a,3}& & \\
    \vdots&\vdots&  &  &  &\ddots\\
    \end{pmatrix},
\end{align}
where we have defined complex bare eigenfrequencies $\tilde\omega_j=i\omega_j +\frac{\kappa_j}{2}$. Diagonalizing this Hamiltonian, we obtain the complex dressed eigenmodes, $\tilde \omega_j'$, yielding out-coupling $Q_\mathrm{e}$ of the waveguide modes dressed by the microwave and LN coupling. We then perform a nonlinear least squares regression, minimizing the cost function Eq.~(\ref{eq:cost_function}) for the fit parameters $x=\{\kappa,\tilde\omega_b,f,g\}$,
\begin{equation}\label{eq:cost_function}
\begin{split}
    \min_x C(x)=\sum_{\omega_a}\sum_{\omega_j'}\big[&\log(Q_\text{e}(\omega_a,\omega_j',x))\\
    &-\log(Q_\text{e}^{\text{data}}(\omega_a,\omega_j'))\big]^2,
\end{split}
\end{equation}

\begin{figure}[]
    \centering
    \includegraphics{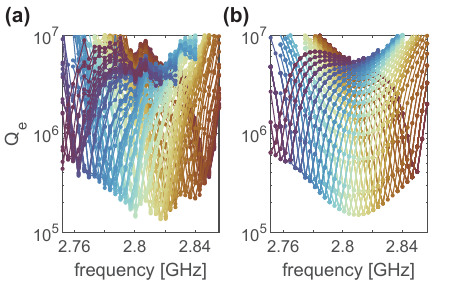}
    \caption{\textbf{Coupling envelope fit}. (a) Measured $Q_\text{e}$ versus dressed mode frequency for different microwave $\omega_c$ (colored curves). (b) Theory $Q_\text{e}$ versus dressed mode frequency generated from fitting the diagonalized Hamiltonian to the data (a).}
    \label{fig:fig_app_QeEnvelope}
\end{figure}

This approach yields values for the fit parameters of, $f/2\pi=\SI{5.63}{\mega\hertz}$, $g/2\pi=1.03\text{MHz}$, $\kappa/2\pi=8.05\text{MHz}$, $\omega_b/2\pi=2.809\text{GHz}$, $\gamma_b/2\pi=12\text{kHz}$. The data and fit $Q_\text{e}$ versus frequency curves are shown in Fig.~\ref{fig:fig_app_QeEnvelope}(a) and Fig.~\ref{fig:fig_app_QeEnvelope}(b) respectively.

\section{Coupling enhancement and frequency noise due to a microwave mode}\label{appendix:coupling_enhancement_microwave_mode}
A system consisting of a transmission line coupled to a dressed mechanical mode, $ b$, with frequency $\omega_b$ and extrinsic (intrinsic) loss rates $\gamma_e$ ($\gamma_i)$, has dynamics that are captured by the following input-output relations,
\begin{align}
    {a}_{\text{out}} = {a}_{\text{in}} + \sqrt{\gamma_e} b,\\
    \frac{\rm d}{\mathrm d t} {b} = - \left(i\omega_b + \frac{\gamma}{2}\right){b} - \sqrt{\gamma_e}{a}_{\text{in}}.
\end{align}
For a frame rotating at $\omega_b$ and in steady state we have,
\begin{align}
    {a}_{\text{out}} = {a}_{\text{in}} + \sqrt{\gamma_e} b,\\
    \left(\frac{\gamma_i+\gamma_e}{2}\right){b}= - \sqrt{\gamma_e}{a}_{\text{in}} \label{eq:one_mode_internal_cavity}.
\end{align}
Solving for ${a}_{\text{out}}/{a}_{\text{in}}$  yields the reflection signal that would be obtained from measuring such a device on resonance,
\begin{equation}
    \frac{{a}_{\text{out}}}{{a}_{\text{in}}} = 1-\frac{2\gamma_e}{ \gamma_e+\gamma_i}.
\end{equation}
In the case that the mechanical mode is very weakly coupled, $\gamma_e\ll \gamma$, the reflection signal from the device is small and in practice is dominated by noise.

If instead, a new mode ${c}$ with frequency $\omega_c$ and extrinsic (intrinsic) loss rates $\kappa_e$ ($\kappa_i)$ is inserted between the transmission line and the mechanical mode and coupled at rate $g$ to the mechanical mode. The input-output relations become,
\begin{align}
    {a}_{\text{out}} = {a}_{\text{in}} + \sqrt{\kappa_e}{c}, \label{eq:two_mode_out_channel}\\
    \frac{\rm d}{\mathrm d t} {c} = - \left(i\Delta + \frac{\kappa}{2}\right){c} - ig{b}  - \sqrt{\kappa_e}{a}_{\text{in}}, \label{eq:two_mode_first_mode}\\
    \frac{\rm d}{\mathrm d t} {b} = - \left(\frac{\gamma_i}{2}\right){b} - ig{c} ,\label{eq:two_mode_second_mode}
\end{align}
where we have chosen to work in the frame rotating at rate $\omega_b$ and have defined $\Delta = \omega_c -\omega_b$. In steady state, these equations are,
\begin{align}
    {a}_{\text{out}} &= {a}_{\text{in}} + \sqrt{\kappa_e}{c},\label{eq:two_mode_ss_out_channel}\\
       \left(i\Delta + \frac{\kappa}{2}\right){c} &=- ig{b}  - \sqrt{\kappa_e}{a}_{\text{in}},\label{eq:two_mode_ss_first_mode} \\
     \left(\frac{\gamma_i}{2}\right){b}&= - ig{c} .\label{eq:two_mode_ss_second_mode}
\end{align}
We solve for $c$, substitute and simplify to obtain an expression for the field $b$, 
\begin{align}
    {a}_{\text{out}} = \left( 1- \frac{\kappa_e}{\left(i\Delta + \frac{\kappa}{2}\right)}\right){a}_{\text{in}} + \frac{- ig \sqrt{\kappa_e}}{\left(i\Delta + \frac{\kappa}{2}\right)}{b}, \label{eq:two_mode_out_channel}\\
     \left(\frac{\gamma_i}{2}+\frac{g^2}{\left(i\Delta + \frac{\kappa}{2}\right)}\right){b}= \frac{ ig\sqrt{\kappa_e}}{\left(i\Delta + \frac{\kappa}{2}\right)}{a}_{\text{in}}. \label{eq:two_mode_internal_cavity}
\end{align}
Inspecting Eq.~(\ref{eq:two_mode_internal_cavity}), we note the similar form to the single mode case Eq.~(\ref{eq:one_mode_internal_cavity}), but with a new term whose imaginary part shifts the frequency $\Delta \omega_b$, and real part modifies the effective extrinsic coupling $\gamma_{e,\text{eff}}/2$,
\begin{align}
    \Delta \omega_b &= \frac{g^2 \Delta}{\Delta^2 + (\kappa/2)^2},\\
    \gamma_{e,\text{eff}} &= \frac{g^2 \kappa}{\Delta^2 + (\kappa/2)^2}.
\end{align}
Note that the effective extrinsic coupling is described by a Lorentzian of width $\kappa$ and peak value $4g^2/\kappa$, and that any frequency noise in the microwave mode is imparted on the mechanical mode through $\Delta \rightarrow \Delta_0 + \delta\omega_c$. In the limit that the two modes are far detuned, and the frequency fluctuations are relatively small, $\Delta_0 \gg \kappa \gg \delta\omega_c$, the mechanical mode fluctuations are $\delta\omega_b \approx \delta \omega_c (g^2/\Delta_0^2)$. This model captures the increased out-coupling as well as excess frequency noise in mechanical modes which may result from the addition of the high-impedance microwave mode.

\section{Coupled mode spectrum fitting}\label{appendix:coupled_mode_fitting}
In this section we detail the approach used to fit to the coupled mode model of the system. The main challenges for fitting the data to the model stem from the large number of fit parameters and the combination of narrow and broad features that must simultaneously be captured by the model. To obtain spectra suitable for fitting, we first take a spectrum with a wide, coarse frequency span that captures the broad, over-coupled microwave mode and allows us to fit the background resulting from cabling, amplifiers, and other microwave elements in the measurement chain. We then take a series of spectra with narrow, fine frequency spans centered around each of the high-Q mechanical modes. We stitch together the coarse and fine datasets, then the fitted background is interpolated to all frequencies and removed from the stitched spectrum. This spectrum can now be fitted to the complex $S_{11}$ signal [Eq.~(\ref{eq:full_S11})]. 

In order to constrain the parameter space and increase the success of nonlinear least squares curve fitting to the spectra, we start by running a particle swarm to roughly optimize a few parameters of the system $x=\{g, f, \omega_{\text{FSR}}\}$ by minimizing the cost function,
\begin{equation}
    \min_x C(x)=\sum_{\omega_c}\sum_{\omega_j}|\omega_j-\omega_j^{\text{data}}|
\end{equation}
which serves to coarsely align the dressed eigenmode frequencies, $\omega_j$, of the diagonalized Hamiltonian. We then run a second particle swarm to perform a more detailed optimization of the bare waveguide mode frequencies using the same cost function, precisely aligning the dressed eigenfrequencies to those in the spectra. Once the dressed frequencies of the model and data agree, we perform a nonlinear least squares curve fit optimization of all system parameters with respect to the cost function.
\begin{align}
\centering
 \min_x C(x)=\sum_{\omega_c} &\sum_i\left[\Re\{S_{11}(\omega_i,x)\}-\Re\{S_{11}^{\text{data}}(\omega_i)\}\right]^2 \nonumber\\
   & +\left[\Im\{S_{11}(\omega_i,x)\}-\Im\{S_{11}^{\text{data}}(\omega_i)\}\right]^2
\end{align}

\section{Experimental Setup}\label{appendix:experimental_setup}
\begin{figure}[h!]
    \centering
    \includegraphics[width=.9\linewidth]{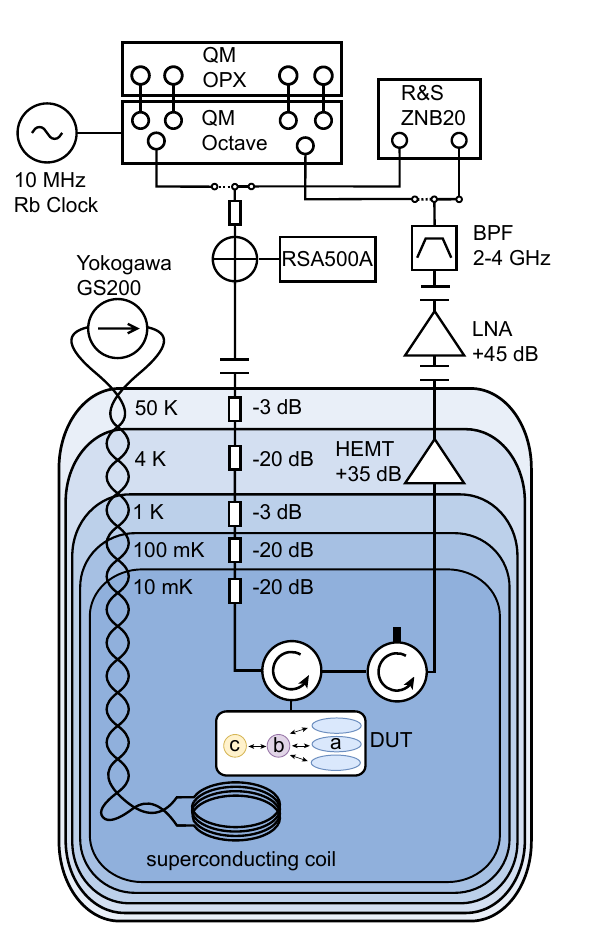}
    \caption{\textbf{Experimental Setup}. \textbf{a}. Measured $Q_\text{e}$ versus dressed mode frequency for different microwave $\omega_a$ (colored curves). \textbf{b}. Theory $Q_\text{e}$ versus dressed mode frequency obtained from fitting the diagonalized Hamiltonian to the data.}
    \label{fig:fig_app_setup}
\end{figure}

An overview of the experimental setup is shown in Fig.~\ref{fig:fig_app_setup}. A Yokogawa GS200 is used as a current source to tune the off-chip superconducting magnetic coil. Cryogenic circulators (Quinstarr QCY-G0250401AM) are used for isolation at the $\SI{10}{\milli\kelvin}$ stage. Cryogenic (Low Noise Factory LNF-LNC0.3\_14A) and room-temperature (Narda-MITEQ LNA-40-00101200-17-10P) low noise amplifiers provide a gain of \SI{35}{\dB} and \SI{40}{\dB} respectively  for the measurement. A band pass filter (Keenlion KBF-2/4-Q7S)is used to filter out any spurious harmonics that may arise due to the amplifiers. The frequency domain measurements are performed with a Vector Network Analyzer (Rohde and Schwartz ZNB20).

The time domain measurements are performed with a Quantum Machines OPX and Octave system. The measurements begin by synthesizing an IF frequency $\approx~\SI{100}{\MHz}$, which is then upconverted to an RF frequency corresponding to the resonances of the device, $\approx~\SI{2.8}{\GHz}$. We synthesize four such tones, each corresponding to a distinct mode, and reflect them off the device for $T_{\text{drive}}=\SI{1}{\ms}$, driving the modes into steady state. After the drive is finished, we allow the mode resonances to ring down while sampling the downconverted reflected field with the OPX ADC. The signal is simultaneously demodulated in $\tau_{\text{demod}}=\SI{8}{\us}$ slices at each of the four drive frequencies, yielding $(I,Q)$ quadrature pairs for each mode as a function of ring down time. These $(I,Q)$ pairs are then combined and averaged in real time to yield the cavity ring down energy, amplitude and equal-time cross-correlation dynamics.

\section{Observable decays for frequency and phase noise processes}\label{appendix:frequency_phase_noise_models}
As discussed in the main text, the ringdown of a mode will be given generally by,
\begin{equation}
    a(t) = a_0 e^{-\kappa t/2} e^{-i\phi(t)},
\end{equation}
where $\phi(t)=\int_0^t \delta\omega(t') dt'$, which can be averaged over many samples of the noise,
\begin{equation}
    \langle a(t)\rangle = a_0 e^{-\kappa t/2} \langle e^{-i\phi(t)}\rangle.
\end{equation}
We focus on the phase term, which we begin by Taylor expanding as,
\begin{equation}
    \langle e^{-i\phi(t)}\rangle = \sum_k\frac{i^k}{k!}\langle \phi(t)^k\rangle.
\end{equation}
To further simplify this expression, we must assume properties of the moments of the noise distribution underlying $\phi(t)$. 

\textit{If $\phi(t)$ is zero-mean}, then all odd moments vanish and we obtain,
\begin{equation}
    \langle e^{-i\phi(t)}\rangle = \sum_k\frac{(-1)^{k}}{(2k)!}\langle \phi(t)^{(2k)}\rangle.
\end{equation}
\textit{If $\phi(t)$ is zero-mean and follows a Gaussian distribution} then all the even higher order moments are $\langle\phi(t)^{(2k)}\rangle=\langle\phi(t)^{2}\rangle^{k}(2k-1)!!$, which leads to the expression,
\begin{equation}
    \langle e^{-i\phi(t)}\rangle = e^{-\langle \phi(t)^2 \rangle/2},
\end{equation}
that depends only on the second moment of the noise process. If the frequency content of the noise process is described by a spectral density $S_\phi(\omega)$, the variance of the phase noise will be given by:
\begin{equation}\label{eq:phi_squared_integral}
    \langle \phi(t)^2\rangle = \int_{-\infty}^{\infty}d\omega S_{\delta\omega}(\omega) F(t,\omega),
\end{equation}
where $F(t)=4\sin^2(\omega t/2)/\omega^2$ is the filter function corresponding to free decay and $S_{\delta\omega}(\omega)$ is the spectral density of frequency noise. 
For white noise (frequency independent spectral density), Eq.~\eqref{eq:phi_squared_integral} evaluates to $\langle \phi(t)^2\rangle=\kappa_\phi t$, with $\kappa_\phi=2\pi S_{\delta\omega}(0)$, causing the field to decay as an exponential linear in time.

If however, the spectral density of the phase noise has a more complicated functional form, the field decay can occur with higher powers of time. 
For example, a Lorentzian spectral density $S_{\delta\omega}(\omega)=A^2\tau/(1+(\omega\tau)^2 )$ arising from an Ornstein-Uhlenbeck process, where $A$ and $\tau$ represent the strength and correlation time of the noise respectively, produces a phase noise variance of:
\begin{equation}
    \langle \phi(t)^2\rangle = \pi A^2\tau^2(t/\tau -1 + e^{-t/\tau}).
\end{equation}
In the regime of fast phase fluctuations $\tau \ll t$, the variance limits to the white noise spectrum previously discussed, scaling linearly in time, $\langle \phi(t)^2 \rangle\sim t$. In contrast, slow phase fluctuations $\tau \gg t$ result in a qualitatively different variance which scales quadratically in time, $\langle \phi(t)^2 \rangle \sim t^2$ \cite{Fanciulli09}.

The functional form of the amplitude damping term can also be solved for random telegraph noise~\cite{Marquardt2009}, which in the limit of fast switching coincides with the Gaussian result, whereas in the slow switching limit produces coherent oscillations which can for short timescales be expanded to produce amplitude damping terms that are quadratic in time.

\section{Phonon number estimation for ringdown measurements}\label{appendix:phonon_number_estimation_for_ringdown_measurements}
To estimate the initial intracavity phonon number for a ringdown measurement, we begin with a spectroscopic measurement of the mode lineshape which we fit to a Lorentzian and extract the corresponding $\kappa_e$. The usual method of calculating the steady-state intracavity phonon number from input-output theory requires the total spectroscopic linewidth, which is not equal to the energy decay rate measured by a ringdown due to the presence of frequency noise. For our ringdown measurement, we obtain both energy and amplitude decay parameters, from which we calculate a $T_2$ time. We invert this to obtain an estimate of the total linewidth $\kappa_{\text{total}}=1/T_2$, which then enters into our estimate of the phonon number:
\begin{equation}
    \langle n \rangle = \frac{P}{\hbar \omega} \times \frac{4\kappa_e}{\kappa_{\text{total}}^2}.
\end{equation}
Here, $P$ is the power in watts incident on the device, $\omega$ is the resonant frequency of the mode.

\section{Phase noise cross correlation}\label{appendix:phase_noise_cross_correlation}
In this section we outline the method used to quantify the cross correlation strength in frequency/phase noise between two modes. The individual mode fields and the correlated field averaged over many realizations of the noise are,
\begin{align}
    \langle a_1(t)\rangle&= a_{1,0}e^{-\kappa_1t/2}   \langle e^{-i\phi_1(t)}\rangle \\
    \langle a_2(t)\rangle &= a_{2,0}e^{-\kappa_2t/2}  \langle e^{-i\phi_2(t)}\rangle \\
    \langle a_1(t) a_2^*(t)\rangle&= a_{1,0}a_{2,0}e^{-(\kappa_1+\kappa_2)t/2} \langle e^{-i(\phi_1(t) - \phi_2(t))}\rangle
\end{align}
if we Taylor expand the dephasing term and assume only that $\phi(t)$ is a zero mean process, then we can rewrite the amplitude decay as,
\begin{align}
\langle e^{-i\phi(t)}\rangle &= 1-\frac{\langle \phi(t)^2 \rangle}{2}+\cdots\\
&=e^{-f(t)}
\end{align}
where $f(t)$ is an arbitrary function that captures the dephasing dynamics. We can also Taylor expand the dephasing term in the correlated field and rewrite it as,
\begin{align}
    \langle e^{-i(\phi_1(t) - \phi_2(t))}\rangle &= 1 - \frac{\langle \phi_1(t)^2 \rangle}{2} - \frac{\langle \phi_2(t)^2 \rangle}{2} \nonumber \\& + \langle\phi_1(t)\phi_2(t)\rangle+\cdots\\
    &=e^{-f_{12}(t)}
\end{align}
where $f_{12}(t)$ captures the correlated dephasing dynamics and is related to the individual dephasings by, 
\begin{align}
    f_{12}(t)=f_1(t)+f_2(t) - (\langle \phi_1(t)\phi_2(t)\rangle + \cdots)
\end{align}
note that if the phase noise is uncorrelated, then the correlated terms in the above expression vanish and the correlated field will dephase consistent with the individual modes, however, the presence of correlated phase noise will serve to reduce the rate of dephasing in the correlated signal, $f_{12}(t) < f_1(t) + f_2(t)$.

To quantify the degree of correlation, we find the time $T_{\phi,{\rm xc}}$ such that
\begin{align}
    \frac{\langle a_1(t) a_2^*(t)\rangle}{\sqrt{\langle |a_1(t)|^2\rangle\langle |a_2(t)|^2\rangle}}=e^{-f_{12}(t)}=e^{-1}
\end{align}
similarly, we find the timescale for the case of uncorrelated, independent phase noise, $T_{\phi,I}$
\begin{align}
    \frac{|\langle a_1(t)\rangle \langle a_2^*(t)\rangle|}{\sqrt{\langle |a_1(t)|^2\rangle\langle |a_2(t)|^2\rangle}}=e^{-f_{1}(t)-f_2(t)}=e^{-1}
\end{align}
by comparing these two timescales, we define the correlation strength metric,
\begin{align}
    \Phi = 1-\frac{T_{\phi,I}}{T_{\phi,{\rm xc}}}
\end{align}
for perfectly correlated dephasing $\Phi\rightarrow1$, for perfectly uncorrelated dephasing $\Phi\rightarrow0$ and for anti-correlated dephasing $\Phi<0$.


\bibliography{main_bib}


\end{document}